\input{aipcheck}

\documentclass[
    ,final            
  ]
  {aipproc}

\layoutstyle{8x11single}

\begin{document}

\title{Quantum control using diabatic and adiabatic transitions}


\author{D.\ A.\ Wisniacki}{
address={Departamento de F\'{\i}sica ``J.\ J.\ Giambiagi", 
Universidad de Buenos Aires, Ciudad Universitaria, Pab.\ I,
C1428EHA Buenos Aires, Argentina}
                        }

\author{G.\ E.\ Murgida,}{
address={Departamento de F\'{\i}sica ``J.\ J.\ Giambiagi",
Universidad de Buenos Aires, Ciudad Universitaria, Pab.\ I,
C1428EHA Buenos Aires, Argentina}
                           }

\author{P.\ I.\ Tamborenea}{
address={Departamento de F\'{\i}sica ``J.\ J.\ Giambiagi",
Universidad de Buenos Aires, Ciudad Universitaria, Pab.\ I,
C1428EHA Buenos Aires, Argentina}
                           }

\begin{abstract}
We exploit the concept of Landau-Zener transitions at avoided energy
crossings as a quantum-control tool.
In an avoided crossing the two quantum states interchange their characteristics
as an external parameter is varied.
Depending on the rate of change of the parameter it is possible to control
the final state.
We use this simple idea to travel along the energy spectrum of a realistic
system: two interacting electrons confined in a quasi-one-dimensional 
semiconductor system.
\end{abstract}

\maketitle


Quantum control of individual electrons in semiconductor nanostructures has 
become an active field of research.
Transport of single electrons through quantum dots has been demonstrated 
in the early 90s \cite{turnstiles} and numerous theoretical studies exploring 
coherent quantum manipulation of one- \cite{1e} and two-electron \cite{2e}
systems have appeared.
Due to the potential applications of this field to quantum information
processing, it is highly desirable to develop simple and versatile
coherent control methods for these semiconductor systems.

\begin{figure}[hbp]
\includegraphics[width=7cm,angle=-90]{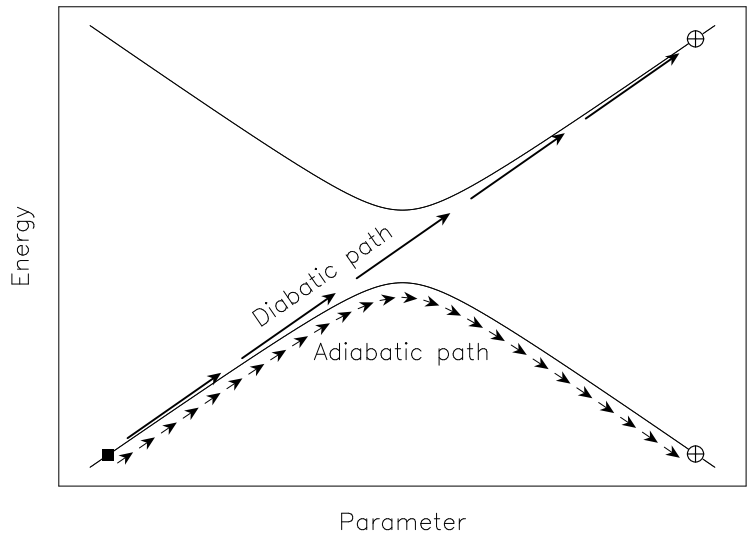} 
\caption{An avoided crossing and a schematic drawing of the diabatic (fast)
and adiabatic (slow) ways to cross it.
Long (short) arrows represent fast (slow) variations of the control parameter.}
\label{fig:avoided_crossing}
\end{figure}

In this presentation we will report our recent work on control of quantum 
systems based on the navigation through the energy spectrum, considered
as a function of an external, controllable parameter \cite{mur-wis-tam}.
The basic element of our control strategy is the avoided crossing.
At an avoided crossing, two energy level approach each other and the
associated eigenstates exchange their characteristics as the external 
parameter sweeps through the crossing.
Zener analyzed this situation in his celebrated paper of 1932 \cite{zen},
he determined the asymptotic transition probabilities as the control
parameter changes linearly with time.
For rapid changes the final state has the same characteristics as the initial
state (diabatic path, see Fig.\ \ref{fig:avoided_crossing}).
Conversely, for adiabatic transitions the initial and final states have 
different characteristics.
Clearly, this entails a very simple control method.
We have shown that this strategy can be used to navigate the spectrum to 
reach any desired target state.

\begin{figure}[hbp]
\includegraphics[width=7cm,angle=-90]{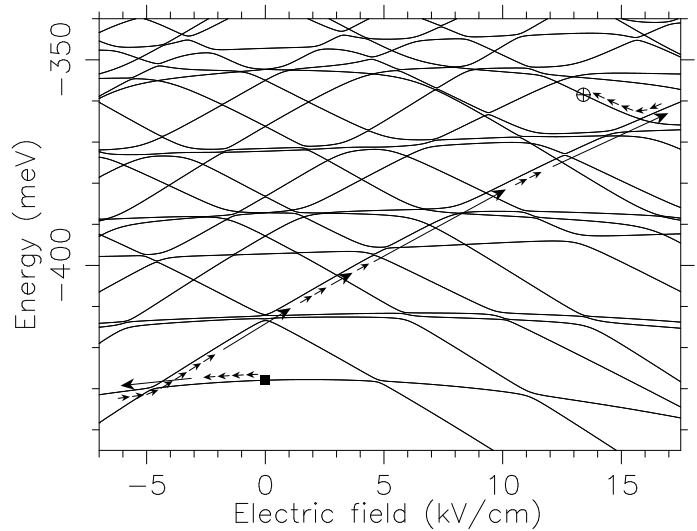}
\caption{The energy spectrum of the two interacting electrons confined in
a quasi-one-dimensional double-well semiconductor nanostructure.
The arrows indicate the intended path to be followed in order to go from
the initial state (filled square) to the target state (circle).
Long (short) arrows represent fast (slow) variations of the 
external electric field used for control.}
\label{fig:navigation}
\end{figure}

\begin{figure}[ht]
\includegraphics[width=10cm,angle=-90]{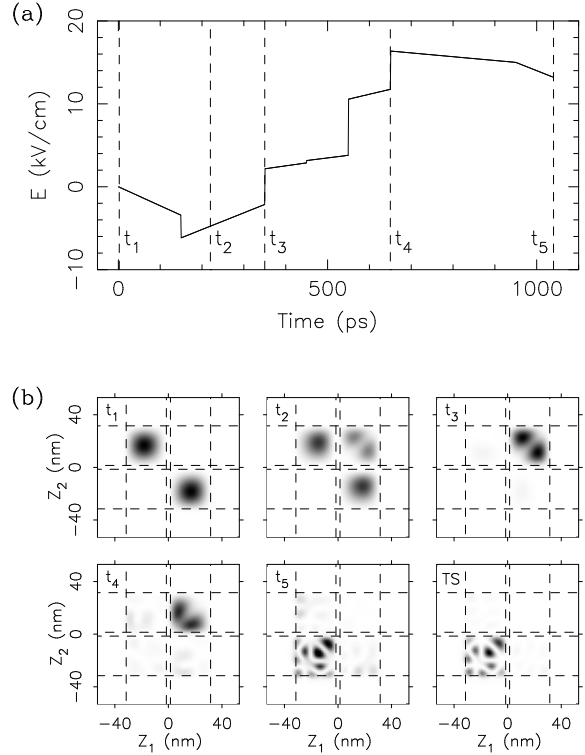}
\caption{(a) The time-dependent applied electric field used to carry our 
the control strategy shown in Fig.\ 2. 
(b) The evolving two-electron wavefunction at different times during
the evolution.  
$t_1=0$, $t_2=276$~ps, $t_3=350$~ps, $t_4=650$~ps, and $t_5=1040$~ps.
Also shown is the target state.}
\label{fig:Efield_and_wavefunctions}
\end{figure}

We apply our control method to a realistic system: a quasi one-dimensional 
double quantum dot. 
The structure contains two interacting electrons and a uniform electric 
field is applied. 
The Hamiltonian of the system is
\begin{equation}
H=-\frac{\hbar^2}{2m}\left(\frac{\partial^2}{\partial z_1^2} +
\frac{\partial^2}{\partial z_2^2}\right) + V(z_1) + V(z_2) + V_C(|z_1-z_2|) -
e(z_1 + z_2)E(t), 
\label{eq:hamiltonian}
\end{equation}
where $z_1$ and $z_2$ are the longitudinal coordinate of the electrons, 
$m$ is the electronic mass, $V_c$ is the Coulomb interaction between the 
electrons, $V$ is the confining potential, and $E(t)$ is an external 
time-dependent electric field.
The confining potential is a double quantum well with well width of 28~nm,
interwell barrier of 4~nm, and 220~meV deep (a typical depth for a
GaAs-AlGaAs quantum well).
Since the Hamiltonian is spin independent, the total spin is conserved
and the spatial wave function remains symmetric under particle exchange
at all times.

We have used the external electric field as the control parameter.
In Fig.\ \ref{fig:navigation} we show the spectrum of eigenenergies of the
Hamiltonian of Eq.\ (\ref{eq:hamiltonian}) versus electric field.
The spectrum is composed of nearly straight lines that never cross, that
is, all level crossings are avoided. 
Far from the avoided crossings, the eigenfunctions have a distinct kind 
of localization depend on the slope of the eigenenergie as a function of 
the electric field \cite{mur-wis-tam}.
For zero slope the electrons are delocalized, that is, each electron is 
in a different dot, for negative slopes both electrons are in the left 
dot and for positive slopes both electrons are in the right dot.

We now describe a very ambitious control example that complements our previous
results of Ref.\ \cite{mur-wis-tam}.
We assume that the ground state of the system without applied electric field 
is the initial state and the desired target state is a high-lying state
(17th state) of the system with an electric field of $E = 13$~kV/cm 
(see Figs.\ \ref{fig:navigation},\ref{fig:Efield_and_wavefunctions}).
The target state belongs to a spectrum line with negative slope, and therefore
has mostly a Righ-Right type of localization (both electrons on the right 
well; see Fig.\  \ref{fig:Efield_and_wavefunctions}b).
The intended path to be travelled is shown in Fig.\ \ref{fig:navigation}.
Short arrows show adiabatic changes of the control parameter and long arrows 
denote diabatic transitions.
The electric field as a function of time is shown in 
Fig.\ \ref{fig:Efield_and_wavefunctions}a;  
the diabatic transitions are seen as vertical jumps (for example, at
$t = 150, 350, 450, 550, 650$~ps) and the straight lines with small slopes
correspond to the adiabatic changes of the electric field.
The evolving wave function at various times is shown in 
Fig.\ \ref{fig:Efield_and_wavefunctions}b; 
the target state is also shown in this figure, labeled as TS.
An important point of our method is that far from the avoided crossings
we move adiabatically and we use either type of evolution (slow or fast) 
at the avoided crossings, depending on which branch of the crossing we wish 
to arrive at.
Furthermore, we use diabatic transitions whenever we need to jump a cluster
of levels. 
See for example in Fig.\ \ref{fig:navigation} the jump from  $E=5$~kV/cm 
to $E = 11$~kV/cm, 
which happens at $t=550$~ps, as seen in 
Fig.\ \ref{fig:Efield_and_wavefunctions}a.
 
We have obtained a similar degree of success for the complex path of 
Fig.\ \ref{fig:navigation} as that obtained in Ref.\ \cite{mur-wis-tam}.
The overlap between the evolved final state (state at $t_5=1040$~ps in 
Fig.\ \ref{fig:Efield_and_wavefunctions}b) 
and the target state (state TS in the same figure) is 0.91.
One can see that the main difference between both states is that the
evolved state includes a some amount right-left probability, while the 
target state has almost none.
On the other hand, the left-left probability distribution for both states, 
which carries most of the weight, is remarkably similar.

In summary, we have developed an efficient method to control the wave 
function of a system dependent on a parameter.
We exploit the diabatic and adiabatic Landau-Zener transitions at
avoided crossings, in order to travel through the energy spectrum.
We have successfully applied our method to a two-electron 
quasi-one-dimensional double-dot semiconductor system.
In this system the control parameter is an external electric field
that interacts with the electrons via the dipole coupling.

\begin{theacknowledgments}
The authors acknowledge support from CONICET (PIP-6137, PIP-5851) and
UBACyT (X248, X179). 
D.A.W.\ and P.I.T.\ are researchers of CONICET.
\end{theacknowledgments}



\begin{thebibliography}{99}

\bibitem{turnstiles}
L.\ J.\ Geerligs {\it et al.},
Phys.\ Rev.\ Lett.\ {\bf 64}, 2691 (1990);
H.\ Pothier {\it et al.},
Europhys.\ Lett.\ {\bf 17}, 249 (1992);
L.\ P.\ Kouwenhoven {\it et al.},
Phys.\ Rev.\ Lett.\ {\bf 67}, 1626 (1991).

\bibitem{1e} See for example
F.\ Grossmann, T.\ Dittrich, P.\ Jung, and P.\ H\"anggi, 
Phys.\ Rev.\ Lett.\ {\bf 67}, 516 (1991)
and  
R.\ Bavli and H.\ Metiu,
Phys.\ Rev.\ Lett.\ {\bf 69}, 1986 (1992).

\bibitem{2e} See for example
P.\ I.\ Tamborenea and H.\ Metiu, 
Phys.\ Rev.\ Lett.\ {\bf 83}, 3912 (1999)
and
P.\ Zhang and X.\-G.\ Zhao,
Phys.\ Lett.\ A {\bf 271}, 419 (2000).

\bibitem{mur-wis-tam} G.\ E.\ Murgida, D.\ A.\ Wisniacki, and P.\ I.\ 
Tamborenea, Phys.\ Rev.\ Lett.\, in press.

\bibitem{zen} C.\ Zener, Proc.\ R.\ Soc.\ London, Ser.\ A {\bf 137}, 696
(1932).

\end{thebibliography}
\end{document}